\documentclass[12pt, amsmath, amssymb]{iopart}
\usepackage{mathrsfs}
\usepackage{graphicx}
\usepackage{amssymb}
\usepackage{epsf,latexsym}
\usepackage{times}
\usepackage{ifpdf}
\ifpdf
\usepackage{epstopdf}
\fi

\newcommand{\ket}[1]{\ensuremath{\vert#1\rangle}}

\newcommand{\beq}{\begin{equation}}
\newcommand{\eeq}{\end{equation}}
\newcommand{\beqa}{\begin{eqnarray}}
\newcommand{\eeqa}{\end{eqnarray}}

\begin{document}

\bibliographystyle{unsrt}

\title{Large Spin Entangled Current from a Passive Device}

\author{Avinash Kolli}
\ead{avinash.kolli@materials.ox.ac.uk}
\address{Department of Materials, University of Oxford, OX1 3PH, U. K.}

\author{Simon C. Benjamin}
\address{Department of Materials, University of Oxford, OX1 3PH, U. K.}
\address{Centre for Quantum Technologies, National University of Singapore, 3 Science Drive 2, Singapore 117543.}

\author{Jose Garcia Coello}
\address{Department of Physics and Astronomy, University College London, WC1E 6BT, U. K.}

\author{Sougato Bose}
\address{Department of Physics and Astronomy, University College London, WC1E 6BT, U. K.}

\author{Brendon W. Lovett}
\ead{brendon.lovett@materials.ox.ac.uk}
\address{Department of Materials, University of Oxford, OX1 3PH, U. K.}

\begin{abstract}
We show that a large entangled current can be produced from a very simple passive device: a cluster of three resonant quantum dots, tunnel coupled to one input lead and two output leads. Through a rapid first order resonant process within the cluster, entangled electrons pairs are emitted into separate leads. We show that the process is remarkably robust to variants in systems parameters. An ideal `clean' mode gives way to a `dirty' mode as we relax system constraints, but even the latter produces useful entanglement. The simplicity and robustness should permit experimental demonstration in the immediate future. Applications include quantum repeaters and unconditionally secure interfaces.
\end{abstract}

\maketitle

\section{Introduction}

Entanglement lies at the heart of quantum information processing. It is an essential resource that must be generated and consumed in the execution of quantum algorithms~\cite{nielsen00}. The ability to generate entanglement between elements that can be well separated spatially would be particularly powerful. In the context of computation, this would allow the linking of small, nanoscale quantum registers~\cite{spiller07}, or the building of cluster states in a distributed architecture~\cite{barrett05, lim05}. Moreover it would enable specific few-qubit functions related to quantum communication; this includes quantum repeaters~\cite{briegel98} for sharing entanglement over arbitrarily long distances, and certain approaches to quantum key distribution. The latter may even enable a secure local interface such as an ATM that is invulnerable to malicious devices interposed between the internal mechanism and the user's identification card~\cite{duligall06}.

There have been a number of recent proposals for generating such entanglement between electron spins that can be subsequently separated by macroscopic distances. For example, two static spins can be entangled by a third passing spin~\cite{costa06,habgood08} or by a sea of conduction electrons~\cite{legel07}. Alternatively, a divided spin chain~\cite{damico07} can be used in which a single excitation spilts into two entangled parts. A further idea is that a current carrying lead can somehow bifurcate, producing pairs of entangled spins that propagate down different leads~\cite{saraga03, oliver02}.

\begin{figure}[t]
\centering
\includegraphics[width=0.6\textwidth]{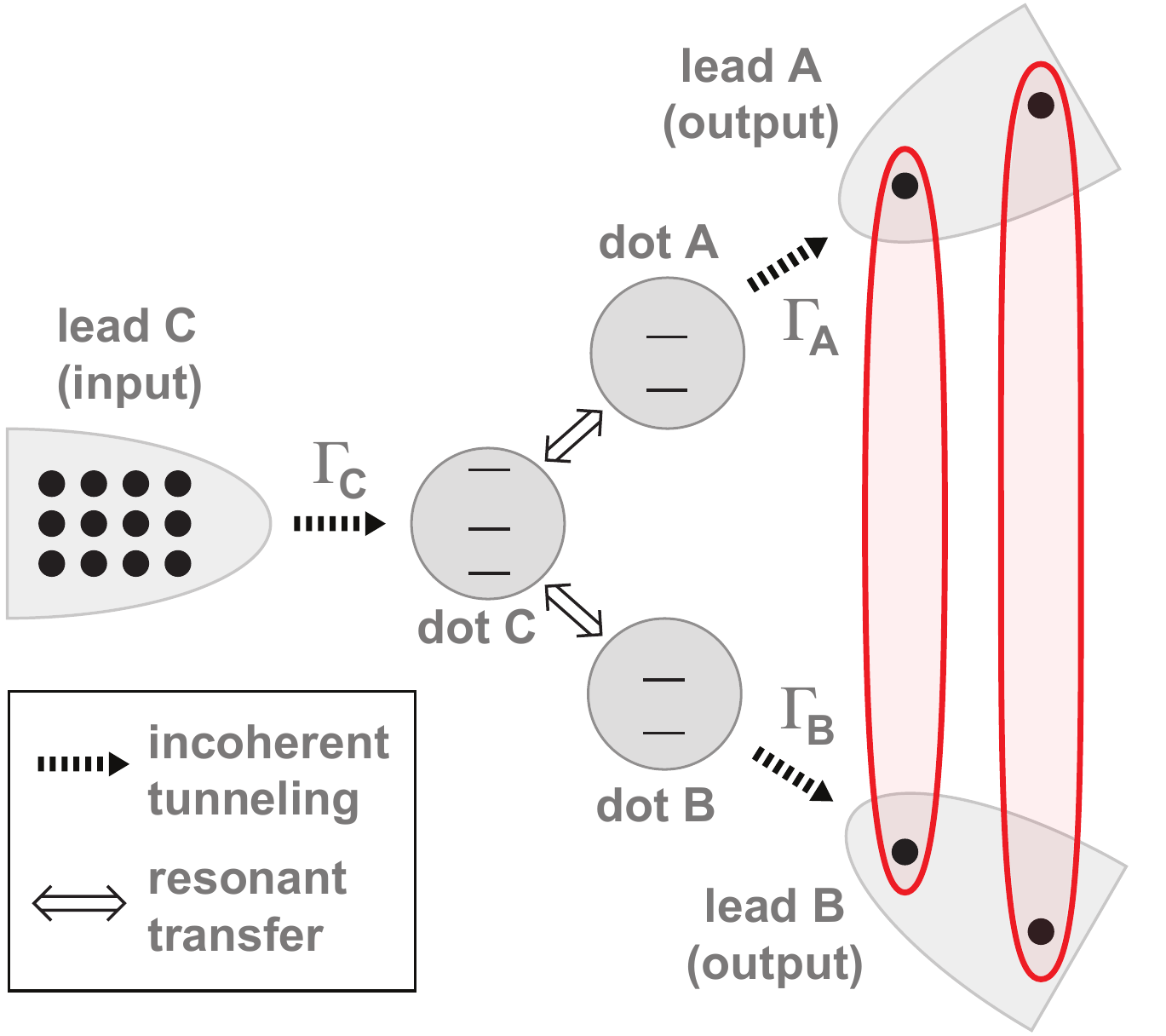}
\caption{The entangler consists of a cluster of three quantum dots, each of the three being coupled to its own lead. Electrons tunnel incoherently from the input lead (left) onto dot $C$, populating a singlet state on that dot. This state is coherently coupled to two other states of the cluster; from these states the electrons can incoherently tunnel to the output leads. For certain parameter regimes, the electrons in each entangled pair will exit to {\em different} leads with high probability.}
\label{fig:model}
\end{figure}

In this paper, we focus on the goal of creating a {\em large} current of spin-entangled electron pairs, from the {\em simplest possible} passive solid state device. Our solution is illustrated in Fig.~\ref{fig:model}; it consists of a single input lead $C$ and two output leads $A$ and $B$, coupled to a core formed of three quantum dots (QDs). The whole device would typically be lithographically defined by gates that can deplete a two dimensional electrons gas~\cite{koppens06, petta05}. Such a structure allows for exquisite control of both electron energies and tunnel couplings. Our device would require an initial configuration, but would then run passively.
In what follows we will show that, under suitable conditions, two electrons enter the QDs and their spins become entangled with each other, and they then leave with this entanglement still intact. Moreover, the entanglement is channelled such that one electron leaves through one output lead, and the other electron leaves through the other lead. This is achieved through a resonant transfer of the entangled electrons, one each to dots $A$ and $B$. We shall show that the entanglement generation and separation process within the QD system occurs on a timescale set directly by the coherent tunneling rate: this is faster than an alternative entangling device discussed in Ref.~\cite{saraga03}, which relies on a second order tunneling process. Our device is fully resonant which allows us to  generate a variety of entanglement qualities at different rates by changing system parameters.

\begin{figure}[t]
\centering
\includegraphics[width=0.8\textwidth]{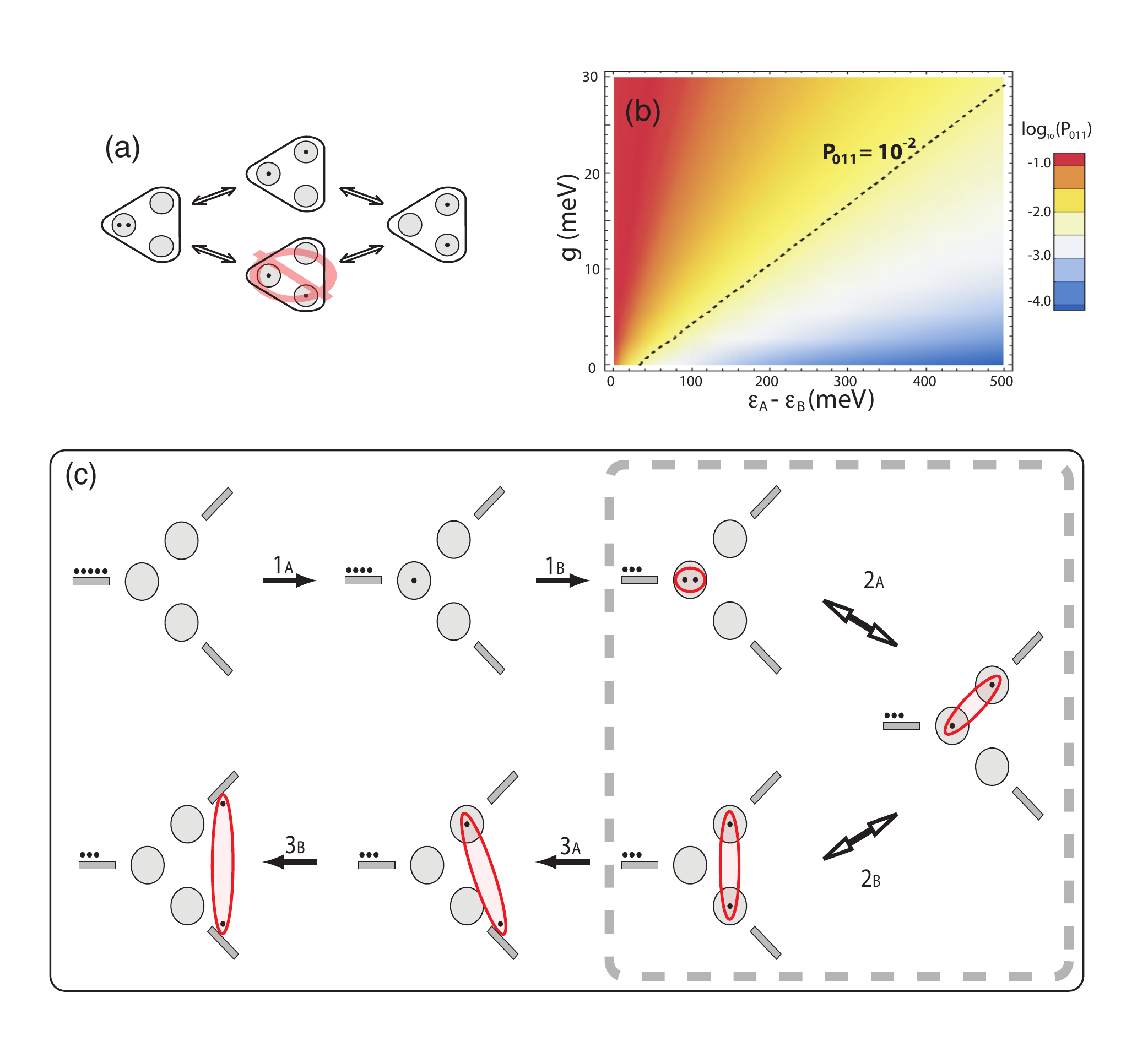}
\caption{(a) The electrons may coherently tunnel from state \ket{002} to \ket{110} via two possible routes; we require that one route is suppressed. (b) Average population within the suppressed state \ket{011} as a function of detuning between $A$ and $B$, $\epsilon_{A}-\epsilon_{B}$, and resonant coupling strength $g$. We see that suppression of state \ket{011} may be achieved for a wide range of the parameter space. (c) Resulting cycle of events; incoherent (and irreversible) tunnelling events are denoted by single-headed arrows, while transitions between resonant states are shown by double-headed arrows.}
\label{fig:scheme}
\end{figure}



We begin with an outline description of the process, then we proceed to an analytic treatment and finally a numerical model of the open dynamics.

\section{The Model System}

Our protocol is schematically depicted in Fig. \ref{fig:scheme}. Suppose the dot cluster is initially uncharged. Dot $C$ receives a electron from its lead; this dot is less tightly confined than $A$ or $B$ so that the lone electron cannot resonantly transfer to those dots. However the lowest two-electron state of dot $C$, i.e. the singlet state, is below the lead potential and therefore a second electron can enter, populating this state. Resonant tunneling between dots is now possible. Double occupancy of dot $A$, or of dot $B$, will not occur because of their tighter confinement; therefore the potential resonant states are $\{|002\rangle,|101\rangle,|011\rangle,|110\rangle\}$ (using the notation $\ket{n_A, n_B, n_C}$). We find that asymmetry between A and B leads to exclusion of $\ket{011}$, and the remaining three-state dynamics leads to the desired emission to separate leads.

The three QDs have an internal Hamiltonian given by:
\beqa
{\cal H}& =&  \sum_{i,\sigma} \epsilon_i n_{i,\sigma} + \sum_{i,j,\sigma,\sigma^{'}} U_{i,j} n_{i,\sigma} n_{j, \sigma^{'}}
\\ &&\nonumber+\sum_{i\neq C,\sigma} g (c_{C,\sigma}^{\dagger} c_{i,\sigma} + c_{i,\sigma}^{\dagger} c_{C,\sigma}),
\eeqa
where $c_{i,\sigma}$ is the annihilation operator for an electron of spin $\sigma$ on dot $i \in \{A, B, C\}$ and $n_{i,\sigma} \equiv c_{i,\sigma}^\dagger c_{i,\sigma}$. The first term represents the single particle energy $\epsilon_i$ for each dot, the second Coulomb repulsion  $U_{i, j}$ between electrons on dots $i$ and $j$ and the third tunneling $g$. The dots are arranged such that tunneling is only significant between dot $C$ and either $A$ or $B$. Furthermore, it is assumed that $U_{CC}$ dominates over all other repulsion terms and for simplicity we may set $U_{CC}=U$, $U_{AC}=U_{BC}=V$ and take $U_{AB}=0$

Each lead $i$ is coupled to the neighboring QD $i$ through incoherent tunneling of magnitude $\Gamma_i$. We bias the device and set the lead chemical potentials $\mu_{i}$ such that transport occurs from lead $C$ to leads $A$ and $B$, which can be described by a super-operators $\mathcal{L}_{i}$ acting on the QD density matrix $\rho$:
\begin{eqnarray}
\mathcal{L}_{A} \rho &=& \Gamma_{A} [c_{A,\sigma} \rho c_{A,\sigma}^{\dagger} - \frac{1}{2}\{c_{A,\sigma}^{\dagger} c_{A,\sigma}, \rho\}], \nonumber \\
\mathcal{L}_{B} \rho &=& \Gamma_{B} [c_{B,\sigma} \rho c_{B,\sigma}^{\dagger} - \frac{1}{2}\{c_{B,\sigma}^{\dagger} c_{B,\sigma}, \rho\}], \nonumber \\
\mathcal{L}_{C} \rho &=& \Gamma_{C} [c_{C,\sigma}^{\dagger} \rho c_{C,\sigma} - \frac{1}{2}\{c_{C,\sigma} c_{C,\sigma}^{\dagger}, \rho\}].
\end{eqnarray}

\noindent {\em Resonant Transfer} - We require that the three states \ket{002}, \ket{101} and \ket{110} are on resonance. This may be achieved if the following conditions are met:
\beqa
\epsilon_{C}-\epsilon_{A} = V-U, \ \ \ \
\epsilon_{C}-\epsilon_{B} = -V.
\label{cond2}
\eeqa

\noindent Meanwhile, to suppress transfer via state \ket{011}, we require that:
\beqa
|(\epsilon_{C} - \epsilon_{B}) + (U - V)| \gg |g|, \nonumber \\
|(\epsilon_{C} - \epsilon_{A}) + V| \gg |g|.
\label{cond3}
\eeqa

From conditions~(\ref{cond2} and \ref{cond3}) we obtain a criterion relating the Coulomb terms: $(U-2V) \gg g$. The resonance condition~(\ref{cond2}) can then be satisfied by careful tuning of the on-site energies such that $\epsilon_{A}-\epsilon_{B} = U-2V$. On-site and Coulomb repulsion energies are typically an order of magnitude larger than coherent tunneling strengths: $U,\epsilon \approx 100~\mu~\mathrm{eV}$, and $g\approx 10~\mu\mathrm{eV}$ \cite{kouwenhoven97,oosterkamp98,fujisawa98}. Thus conditions~(\ref{cond2} and \ref{cond3}) can easily be met experimentally. Therefore the essential coherent dynamics are governed by a simple Hamiltonian:
\beq
H^{'} = g (\sqrt{2} c^{\dagger}_{2C} c_{A} (1 - c^{\dagger}_{B} c_{B}) + c^{\dagger}_{C} c_{B} c^{\dagger}_{A} c_{A} + h.c.)
\eeq
where operators $c^{\dagger}_{C} = |1\rangle_{C}\langle 0|$ and $c^{\dagger}_{2C} = |2\rangle_{C}\langle 1|$.

Electrons must tunnel off the device (step 3) on a timescale that is fast compared with the refill time $\Gamma_C^{-1}$ of the system to avoid another electron entering the dot region before the first two leave, which may destroy the created entangled state.  Thus $\Gamma_B, \Gamma_A \gg \Gamma_C$. Ideally, one of the pair of electrons should tunnel into lead $B$ before lead $A$; this prevents the remaining electron from being `stuck' on dot $C$ (i.e. $\Gamma_B \gg \Gamma_A$). However, in what follows we shall see that it is possible to relax this condition.

\section{Monte Carlo Simulations}

In order to completely characterise the performance of the device for real parameters, we now exploit a quantum trajectories formalism to describe the electron dynamics:
\beqa
\rho(t+dt) &=& -i [H^{'},\rho(t)] dt + \sum_{i\in \{A,B\}} (Tr\{\mathcal{J}_{i}\rho\}\rho - \mathcal{A}_{i}\rho ) dt \nonumber\\ &+& \sum_{i\in\{A,B\}} \left(\frac{\mathcal{J}_{i}\rho}{Tr\{\mathcal{J}_{i}\rho\}}\right) dN_{i}(t) + \mathcal{L}_{C}\rho dt + \rho(t) \nonumber\\
\label{qtraj}
\eeqa
where $\mathcal{J}_{i} \rho = c_{i} \rho c^{\dagger}_{i}$ represents the jump super-operator corresponding to an event where an electron hops off dot $i\in\{A, B\}$. $\mathcal{A}_{i} = \frac{1}{2} \{c^{\dagger}_{i} c_{i} , \rho \}$ and $dN_{i}(t)$ is the stochastic increment taking the values $\{0,1\}$, which denotes the number of electrons that hop off dot $i$ in the time interval $t,t+dt$.

\begin{figure}
\centering
\includegraphics[width=0.8\columnwidth]{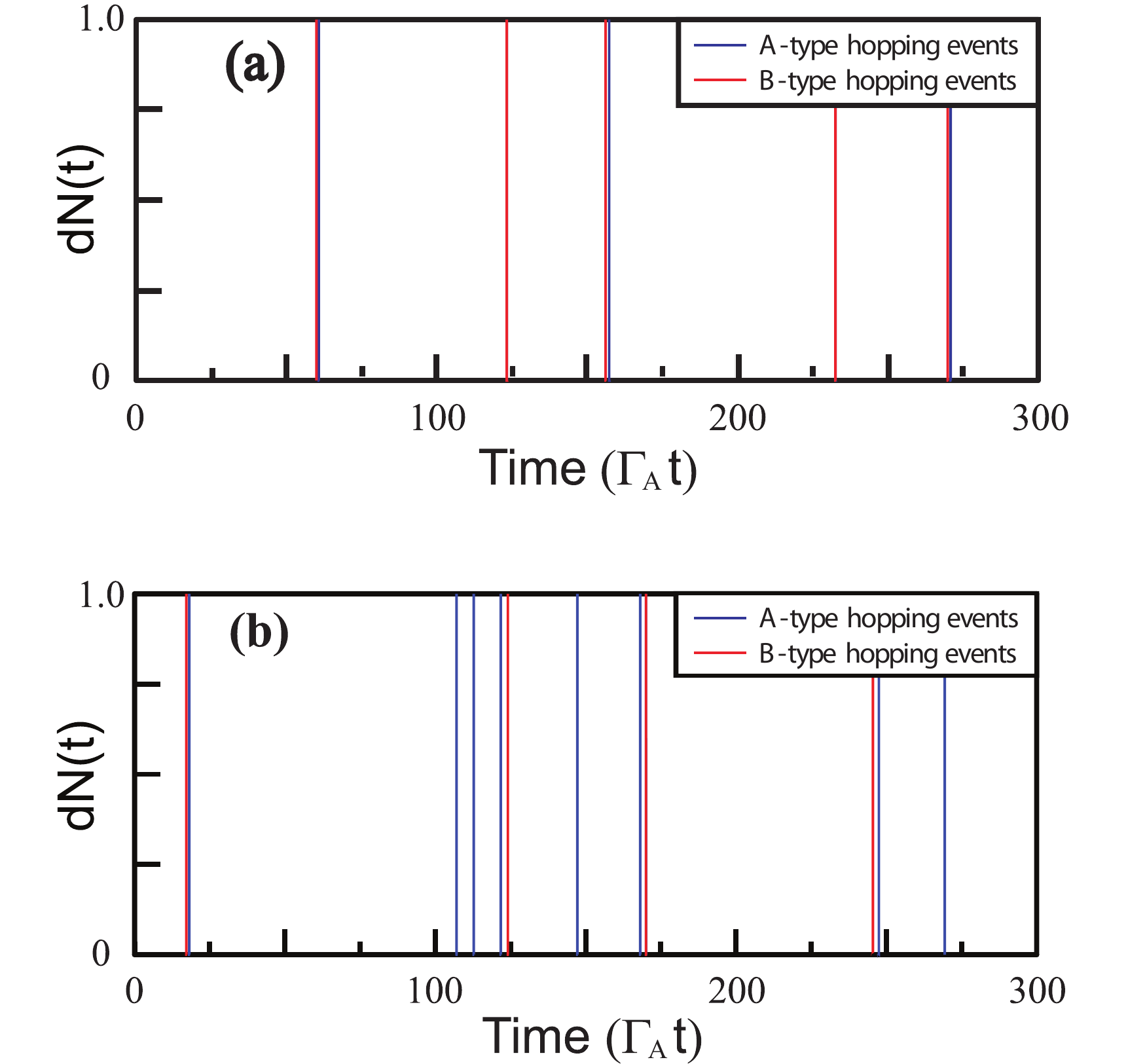}
\caption{Plot hopping events from dot $A$/$B$ to lead $A$/$B$ as a function of time, from typical runs of the simulation. Two regimes are shown: (a){\it clean}: $\Gamma_{B} =10 \Gamma_{A}$, $\Gamma_{C} = \Gamma_A/25$ and $g=10 \Gamma_A$; (b) {\it dirty}: $\Gamma_{B} = \Gamma_{A}$, $\Gamma_{C} = \Gamma_B/25$ and $g=10\Gamma_A$. Within the clean trace, the second and fourth line corresponds to a pair of hopping events, from $B$ then $A$, whose separation cannot be resolved on this plot.}
\label{fig:hoppingevents}
\end{figure}

We use a Monte Carlo method to generate values of the stochastic increment $dN_i (t)$: at every time step two random numbers $r_i$ are generated, and if $r_{i} < \mathrm{Tr}\{\mathcal{J}_{i}\rho\}dt$ then the stochastic increment takes the value $dN_{i}(t) = 1$. We then generate simulations of individual hopping events in two scenarios. The first, which we shall call {\it clean}, is depicted in Fig.~\ref{fig:hoppingevents}(a) and corresponds to the regime where $\Gamma_B \gg \Gamma_A$. In this case, electrons hop onto the two leads in pairs, which from our previous argument are spin entangled with one another. The pairs are clearly grouped together, with typical pair-pair separation $2\Gamma_C^{-1}$. The second regime, which we shall call {\it dirty} is shown in Fig.~\ref{fig:hoppingevents}(b) and corresponds to $\Gamma_B = \Gamma_A$. This simulation includes both pairs and single events because it is now possible for an electron to enter the system before both electrons from the previous cycle have left. However, the relaxed parameter contraints in this latter case may prove more experimentally accessible and so comparing the clean and dirty regime allows us to explore the robustness of our protocol.

\section{Post-Selection}

When two electrons leave the system that are closely separated in time they are almost always emitted into different leads, even in the {\it dirty} scenario. Such electrons have a high probability of being entangled with one another, moreover one could further post-select such `good' pairs by choosing only those electrons whose temporal separation is below some threshold.
Any device of the kind we are describing emits electrons probabilistically and so must have some kind of electron detection system `down stream' of the entanglement generator to collect the pairs prior to subsequent processing. Here we can exploit that necessary detection system as a kind of filter, allowing us to identify those pairs that are most likely to be `good' Bell pairs. We emphasize that this does not involve any additional complexity beyond that which must be present in any case.

Post-selection necessarily involves a decrease in the rate of production of acceptable pairs. We may calculate this rate by defining correlation functions:
\beq
C_{i,j}(\Delta) = \frac{\langle \mathcal{J}_{j}(\Delta) \mathcal{J}_{i}(0)\rangle}{\langle \mathcal{J}_{j}(0)\rangle \langle \mathcal{J}_{i}(0)\rangle},
\eeq
which describe the probability of an electron hopping  to lead $j$ at time $\Delta$, given that an electron hopped to the other lead $i$ at time zero, with no events in the interim. The probability that two electrons separated by a time $\tau$ are `good' is given by $C_{A,B} + C_{B,A}$, and the effective rate of production of such `good' pairs is given by
\beq
\mathcal{R}(\tau) = \frac{\Gamma_{C}}{2} \int_{0}^{\tau} ( C_{A,B}(\Delta) + C_{B,A}(\Delta) ) d\Delta.
\eeq

\begin{figure}
\centering
\includegraphics[width=0.9\columnwidth]{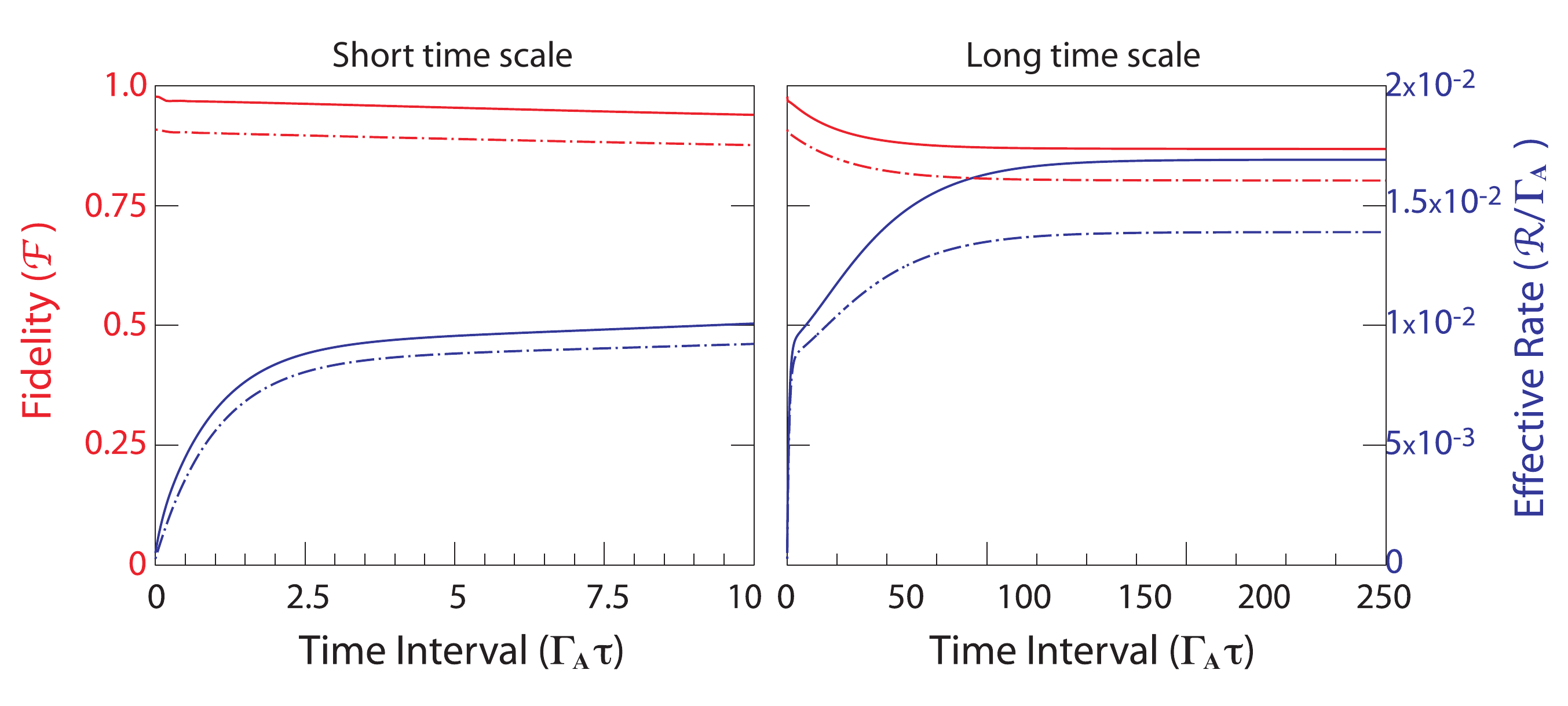}
\caption{Effective rate of production of `good' pairs of electrons (blue) and their entanglement fidelity (red) for short and long time scales in the two parameter regimes {\it clean} (solid) and {\it dirty} (dashed).}
\label{fidrates}
\end{figure}

The effective rates for the {\it clean} and {\it dirty} regimes are plotted in Fig \ref{fidrates}. As the selection time interval increases, the rate of `good' pair production also increases, for both regimes.
In both cases there is a short initial period where the effective rates increase rapidly: this is a signal from electrons already in the system which typically tunnel off at the rate $\Gamma_A$. There follows a longer period where the curves rise much more slowly: this is associated with the time needed for new electrons to enter the system, which is typically $\Gamma_C^{-1}$.
As expected, we find that the maximum rate is larger for the {\it clean} regime than for the {\it dirty} regime.

The rate of pair production is meaningless unless we have a measure of how entangled those pairs are. This can be determined by the probability that the sequence of events depicted in Fig.~\ref{fig:scheme}, which definitely produces an entangled state, has occurred. This probability is:
\beq
p(\Delta) = \int_{0}^{\infty} p_{A,B}(t,t+\Delta) + p_{B,A}(t,t+\Delta) dt
\eeq
where
\beq
p_{i,j}(t,t+\Delta) = \Gamma_{i} \Gamma_{j} \langle (1-n_{i}(t+\Delta)) \mathcal{J}_{j}(t+\Delta) n_{j}(t) \mathcal{J}_{i}(t) \rangle.
\eeq
The initial state $\rho_{0}=(1-n_{A})(1-n_{B}) \rho_{SS}$ is the solution for the system in a steady state, projected onto the subspace with no electrons in either dot $A$ or dot $B$. The probability of definitely obtaining an entangled pair whose separation in time is less than $\tau$ is then
\beq
{\cal P}(\tau) = \frac{\int_0^{\tau} p(\Delta) d\Delta }{\int_0^{\tau} q(\Delta) d\Delta }
\eeq
where
\beq
q(\Delta) = \int_{0}^{\infty} \Gamma_{A}\Gamma_{B} (\langle \mathcal{J}_{A}(t+\Delta) \mathcal{J}_{B}(t)\rangle + \langle \mathcal{J}_{B}(t+\Delta) \mathcal{J}_{A}(t)\rangle) dt.
\eeq
If the sequence in Fig.~\ref{fig:scheme} does not occur, the electrons' state will be a mixture of all possible Bell  states, one quarter of which are the desired states. Thus the fidelity ${\cal F}$ of creating a perfect pair is ${\cal F} = (1+3{\cal P})/4$.
Fig.~\ref{fidrates} shows this fidelity as a function of time in the clean and dirty regimes. As expected, the fidelity of pairs produced drops significantly once the cut-off time is of order $\Gamma_C^{-1}$.


A sensible strategy would be to choose a $\tau$ that's at the end of the fast initial rate rise. At this point, dirty pairs are emitted at $82\%$ the rate as clean pairs, while the fidelity of each dirty pair is only $0.90$, compared with $0.94$ for clean pairs. Both these values are within range where we may use entanglement distillation protocols to generate higher quality entanglement \cite{VolbrechtVerstraete05}. We therefore see that the relaxed, dirty regime performs comparably with the clean regime under these conditions.


We have also considered situations where the various intrinsic coherent coupling strengths within the tri-dot structure are no longer matched. Provided that the hierarchy of rates is respected, the operation of the device is found to be qualitatively identical, and quantitatively very similar, for all such choices.

\section{Decoherence}

In this final section we shall briefly address the effects of decoherence on the device. We must ensure that the operation time of the device is fast compared to the decoherence times of the spin and charge degrees of freedom. Spin decoherence is only an issue once the singlet has been formed on dot C. Therefore the critical timescale for the spins is the time taken for both electrons to leave the tri-dot structure set by $\Gamma_{A}$. From our previous discussions of system parameters we should expect a tunneling rate of $\Gamma_{A}=0.1~\mu\mathrm{eV}$ corresponding to tunneling time of $15~\mathrm{ns}$. Experiments have reported decoherence times of $1.2\mu s$ with refocusing techniques \cite{petta05}. Nuclear spin silent materials would eliminate the need for refocusing. Meanwhile for the charge degrees of freedom, coherence must be maintained on a timescale longer than the time taken for the resonant coherent processes $g$. This can be readily satisfied as $g\approx 10~\mu\mathrm{eV}$ corresponding to a time of $150~\mathrm{ps}$, while charge decoherence times of $1~\mathrm{ns}$ have commonly been observed \cite{petta04,hayashi03}.

\section{Summary}

To summarize, we have proposed a passive device that can produce a stream of spin entangled electrons. We have shown that the rate and fidelity of these electrons depends on the choice of system parameters and can be altered by a degree of post selection. This allows control over the characteristics of the generated entangled electrons that can be chosen to suit the application.
This process is very fast: in practice the overall performance of the device will limited only by rates of incoherent tunnelling and the electronics used to subsequently handle the entangled pairs.

We thank the QIPIRC (No. GR/S82176/01) for support. This work is also supported by the National Research Foundation and Ministry of Education, Singapore. B. W. L. and S. C. B. thank the Royal Society for University Research Fellowships. S. B. thanks the EPSRC Advanced Research Fellowship, the Royal Society and the Wolfson Foundation.

\end{document}